# Security Evaluation of Mobile Banking Applications in Sudan

**Abdelmonim Naway**

**muniem@proton.me**

## Abstract

The rapid digitalization of the Sudanese financial sector has precipitated a surge in Mobile Banking Applications (MBAs); however, this growth has frequently outpaced rigorous security auditing. This study provides a comprehensive technical audit of the four most widely used Sudanese MBAs—Bankak, Fawry, Okash, and Sahil—collectively serving a user base of over 1.6 million. Utilizing Static Application Security Testing (SAST) via the Mobile Security Framework (MobSF) and Quixxi, the applications were evaluated against the OWASP Mobile Application Security Verification Standard (MASVS). Findings were mapped to Common Weakness Enumeration (CWE) identifiers to identify systemic vulnerabilities.

Analysis revealed critical disparities in security posture. Bankak, the market leader, exhibited the highest risk profile (12 vulnerabilities), including a critical absence of SSL certificate pinning and unsafe TrustManager implementations, rendering it highly susceptible to Man-in-the-Middle (MitM) attacks. While Fawry demonstrated relative maturity (7 vulnerabilities), a universal failure was observed across all four applications regarding secure random number generation (CWE-330), potentially compromising session token integrity. Additionally, Bankak and Okash were found to utilize deprecated cryptographic algorithms (MD5/SHA-1). Notably, all applications successfully disabled ADB backups, yet 100% retained verbose debugging symbols in production APKs, significantly lowering the barrier for reverse engineering. This research addresses a critical gap in the national fintech ecosystem by providing actionable technical recommendations for developers and a strategic roadmap for implementing "security-by-design" principles across the sector.

## 1. Introduction

The proliferation of mobile technology has fundamentally restructured the financial services sector. Mobile Banking Applications (MBAs) have transitioned from luxury services to critical infrastructure, handling billions of transactions and sensitive Personally Identifiable Information (PII). However, the Android ecosystem's inherent openness and fragmentation introduce significant security challenges, rendering MBAs primary targets for sophisticated cyber-attacks, including Man-in-the-Middle (MitM) attacks, code injection, and unauthorized data exfiltration.

While global financial institutions adhere to rigorous security frameworks, the security posture of MBAs in emerging economies, particularly in Sudan, remains under-documented. The Central Bank of Sudan (CBOS) reports 37 active banks, most of which have deployed Android-based MBAs to facilitate cashless transactions. Despite this growth, there is a notable scarcity of peer-reviewed empirical research concerning the technical vulnerabilities inherent in these local applications. This lack of scrutiny poses a systemic risk to the Sudanese financial ecosystem, as vulnerabilities in data storage, cryptographic implementations, and network communication can compromise user trust and financial stability.

Existing literature indicates that even high-traffic banking apps globally often suffer from insecure data storage and weak transport layer security. These findings highlight the need to evaluate similar vulnerabilities in the Sudanese market, where empirical data is currently lacking. This study addresses this gap by performing a systematic security evaluation of the four most frequently downloaded Sudanese MBAs from the Google Play Store, utilizing a Static Application Security Testing (SAST) methodology aligned with the OWASP Mobile Application Security Verification Standard (MASVS).

**Research Questions (RQs):**

- **RQ1:** To what extent do leading Sudanese MBAs comply with international security standards regarding data storage, cryptography, and network communication?
- **RQ2:** What are the potential exploitation vectors of the identified vulnerabilities and their projected impact on user data integrity?

**Key Contributions:**

- **Standardized Security Baseline:** Establishes the first forensic-grade security baseline for the Sudanese mobile banking sector using the OWASP MASVS framework.
- **Systemic Vulnerability Identification:** Identifies critical "high-risk" flaws in transport-layer security and data-at-rest encryption.
- **Comparative Analysis:** Provides a cross-institutional audit identifying architectural choices that influence security outcomes.
- **Actionable Remediation Roadmap:** Offers tiered technical and regulatory recommendations for the financial sector.

## 2. Related Studies

The security landscape of MBAs is a multi-dimensional domain involving client-side security, transport layer integrity, and server-side protection. This section categorizes recent scholarly efforts into three primary areas: global vulnerability trends, regional assessments in emerging markets, and the specific research landscape in Sudan.

**2.1 Global and Architectural Vulnerability Assessments** Scholars have long utilized automated analysis to identify systemic risks. He et al. (2015) categorized primary threats, identifying malware, third-party library risks, and phishing as dominant vectors. Structurally, research has evolved toward multi-layered threat modeling. Yildirim & Varol (2019) and Bojjagani & Sastry (2016) proposed hierarchical models evaluating MBAs at the code, communication, and device levels. These frameworks underscore that a comprehensive audit must consider the entire transaction ecosystem beyond the application code.

**2.2 Security Evaluations in Emerging Markets** In developing economies, the adoption of MBAs often outpaces the implementation of security auditing. In the Middle East and Africa, comparative studies have revealed significant gaps in compliance with international standards. Al-Delayel (2022) identified critical vulnerabilities in Qatari MBAs regarding cryptographic implementation. Similarly, Latifa et al. (2017) found that 45% of apps in Morocco, Tunisia, and Algeria requested "risky" permissions without functional necessity. Bassolé et al. (2020) analyzed 53 African MBAs, finding that 71.7% lacked proper access controls over external storage.

**2.3 The Sudanese Context and Research Gap** While mobile banking is critical to Sudan's financial inclusion strategy, technical security research remains embryonic. Most existing literature focuses on adoption factors or user perception. A notable exception is Khalid (2020), which analyzed 13 Sudanese MBAs; however, that study utilized broad criteria with loosely defined selection parameters. The current state of research reveals a significant gap between rapid fintech adoption and the empirical validation of its security infrastructure. This study addresses this systemic gap by introducing a rigorous, standardized methodology based on the OWASP MASVS, transitioning the local discourse from general observation to a systematic audit of critical vulnerabilities.

## 3. Materials & Methods

This study employs a multi-phase security evaluation framework to assess the integrity of the Sudanese mobile banking ecosystem. As illustrated in **Figure 1**, the research follows a structured five-phase methodology: (1) application acquisition and integrity verification; (2) automated static analysis; (3) manifest and resource inspection; (4) mapping findings to international standards; and (5) risk impact synthesis.

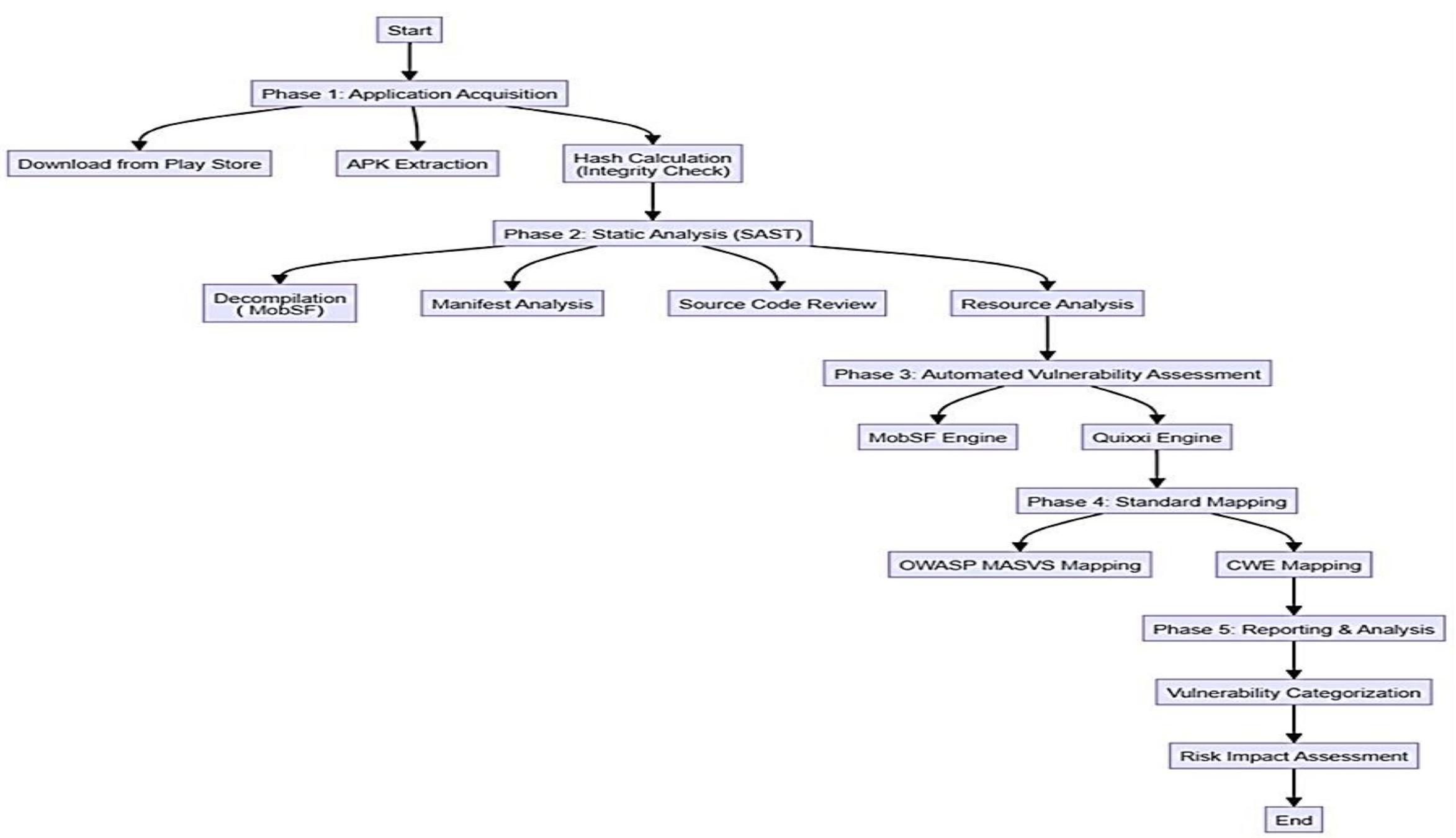

**Figure1: The systematic static evaluation process**

The workflow initiated with the acquisition of target APKs and their integrity verification via SHA-1 and MD5 hashing to ensure a non-repudiated baseline. This was followed by a dual-track static analysis involving automated scanning and manual code review to identify non-compliance with the OWASP MASVS framework.

### 3.1 App Selection and Scope

The evaluated applications were selected based on download frequency and market penetration. **Table 1** details the scope of the evaluated Sudanese mobile banking applications.

| Bank Name | App Alias | Version | Downloads | MD5 Hash | SHA-1 Hash |
| --- | --- | --- | --- | --- | --- |
| **Bank of Khartoum** | Bankak | 4.42 | 1,000,000+ | 55ef30c5419cf2b16e73de82aa0d4325 | 8b38c1f2d08470a0620cd30a62f3201accca4cec |
| **Faisal Islamic Bank** | Fawry | 2.3 | 500,000+ | 3942b642e98b2a107858bcc401b62a17 | f597deecb868e05e2e32b1f16b813687653226e8 |
| **Omdurman National Bank** | Okash | 3.2 | 100,000+ | 6aea996d1f2f207fc3c0691216df27eb | bf31bc1b625ce0811a1ade1c1e5b10b5850bf3aa |
| **Bank Elneil** | Sahil | 1.8 | 50,000+ | af550a2417bc72e6399eef31cc7ea027 | 807e2eec84fbcd831be1e884f5b14089079f0d44 |

**Table 1: Scope of Evaluated Sudanese Mobile Banking Applications**

### 3.2 Security Evaluation Framework

Evaluation criteria were synthesized from OWASP MASVS v2.0, cross-referenced with Common Weakness Enumeration (CWE) identifiers. The assessment categorized vulnerabilities into five critical domains: (1) Data Storage and Privacy; (2) Cryptography; (3) Network Communication; (4) Platform Interaction; and (5) Code Quality. The specific security requirements, their descriptions, and the corresponding MASVS/CWE mappings used as benchmarks for this audit are detailed in Table 2.

**Table 2: Security Evaluation Criteria & Mapping**

| Category | Vulnerability / Criterion | CWE ID | OWASP MASVS v2.0 | Potential Impact / Exploit |
| --- | --- | --- | --- | --- |
| **1. Data Storage and Privacy** | Unsafe file deletion | CWE-200 | MASVS-STORAGE-1 | Information disclosure of sensitive transaction logs |
| | External Storage Access | CWE-276 | MASVS-STORAGE-2 | Incorrect default permissions allow other apps to read data |
| | Missing Copy/Paste Protection | CWE-200 | MASVS-STORAGE-5 | Sensitive data exposure via the system clipboard |
| | Missing Screen Protection | CWE-200 | MASVS-STORAGE-6 | Visual data leakage through screenshots or screen sharing |

| Category | Vulnerability / Criterion | CWE ID | OWASP MASVS v2.0 | Potential Impact / Exploit |
| --- | --- | --- | --- | --- |
| | No Background Blurring | CWE-200 | MASVS-STORAGE-6 | Exposure of sensitive app snapshots in the recent apps menu |
| | Cleartext in Source Code | CWE-312 | MASVS-STORAGE-3 | Static credential theft by reverse-engineering the APK |
| | ADB Backup Allowed | CWE-200 | MASVS-STORAGE-4 | Unauthorized local backup of private application data |
| | Hidden WebView Elements | CWE-919 | MASVS-PLATFORM-2 | Local escalation of privilege through hidden interface components |
| **2. Cryptography** | Weak Hashing (MD5/SHA1) | CWE-327 | MASVS-CRYPTO-1 | Inadequate encryption strength leading to collision attacks |
| | Weak Random Number Gen | CWE-330 | MASVS-CRYPTO-2 | Predictable tokens or keys due to low-entropy algorithms |
| | Predictable IV (CBC Mode) | CWE-329 | MASVS-CRYPTO-2 | Ciphertext manipulation via initialization vector guessing |
| | Insecure Protocols | CWE-327 | MASVS-CRYPTO-3 | Use of broken or risky cryptographic protocols (e.g., TLS 1.0) |
| | Missing SQLite Key Protection | CWE-798 | MASVS-CRYPTO-1 | Access to local databases via hardcoded cryptographic keys |
| **3. Network Communication** | Deprecated SSL Sockets | CWE-310 | MASVS-NETWORK-1 | Susceptibility to Man-in-the-Middle (MITM) attacks |
| | Missing Certificate Pinning | CWE-295 | MASVS-NETWORK-2 | Failure to validate server identity against forged certificates |
| | Unsafe TrustManager | CWE-295 | MASVS-NETWORK-2 | Improper certificate validation allows traffic interception |
| **4. Platform Interactions** | Insecure Activity Export | CWE-926 | MASVS-PLATFORM-1 | Unauthorized launching of internal application screens |
| | Insecure Service Export | CWE-926 | MASVS-PLATFORM-1 | Malicious binding to background services to steal data |
| | Insecure Broadcast Receiver | CWE-927 | MASVS-PLATFORM-1 | Interception of system-wide intents and data broadcasts |
| | Insecure Content Providers | CWE-926 | MASVS-PLATFORM-1 | Unauthorized access to internal application databases |
| | WebView File Access | CWE-919 | MASVS-PLATFORM-2 | Exfiltration of local files through malicious web scripts |

| Category | Vulnerability / Criterion | CWE ID | OWASP MASVS v2.0 | Potential Impact / Exploit |
|---|---|---|---|---|
| **5. Code Quality** | Debugging Info Provision | CWE-215 | MASVS-RESILIENCE-1 | Leakage of system logs and stack traces to attackers |
| | Debuggable App | CWE-215 | MASVS-RESILIENCE-1 | Ability for attackers to attach debuggers and dump memory |
| | Improper SSL Error Handling | CWE-701 | MASVS-NETWORK-3 | Bypassing SSL warnings leading to unencrypted data transit |

### 3.3 Tool Selection and Experimental Setup

**3.3.1 Rationale for Analysis Tools** The efficacy of Static Application Security Testing (SAST) depends on the recency of vulnerability signatures and the ability to parse modern Android architecture. This study utilized **MobSF v3.6.3** as the primary framework for manifest and code-level auditing. To supplement these findings, the **Quixxi Security** platform was employed for automated static analysis. By subjecting the APK files to two independent scanning engines, the study ensures a comprehensive evaluation against the OWASP MASVS framework, allowing for cross-verification of identified security flaws.

**3.3.2 Experimental Environment** Analysis was conducted within a controlled virtualized environment to ensure process isolation. The hardware and software specifications include a Linux-based workstation (Ubuntu 22.04 LTS) to prevent environmental interference during the decompilation and analysis phases. The specific hardware and software specifications, including the virtualization layer and analysis tools, are detailed in Table 3.

**Table 3: Software & Environment Specifications**

| Component | Specification / Tool | Purpose |
|---|---|---|
| **Host OS** | Windows 10 | Primary workstation |
| **Virtualization** | Oracle VirtualBox 6.2.2 | Environment isolation |
| **Guest OS** | Ubuntu 22.04. | Security testing platform |
| **Primary Tool** | MobSF (v3.6.3) | Static Analysis |
| **Secondary Tool** | Quixxi Online | Automated reporting & verification |

## 4. Results and Analysis

This section presents the findings of the static security audit for the four selected Sudanese MBAs. Results are categorized according to the five security domains established in the methodology. For clarity, a cross symbol (☒) denotes an identified vulnerability (failure to meet the standard), while a checkmark (☑) indicates compliance.

### 4.1 Data Storage and Privacy

Static analysis revealed significant deviations from OWASP MASVS-STORAGE requirements. As illustrated in Table 4, the Fawry application demonstrated the highest compliance, meeting five of the eight criteria. Conversely, Okash exhibited the most critical gaps, failing six criteria.

A prevalent issue across all four applications is the lack of protection against copy-and-paste operations from EditText fields and the absence of screenshot restrictions (with the exception of Fawry and Sahil). Notably, all applications successfully disabled ADB backups, preventing unauthorized local data extraction via USB debugging. However, the storage of files in external directories without strict access controls remains a systemic risk, particularly for Okash and Sahil, which failed criteria regarding unsafe file deletion and external storage access. However, it is worth noting that all applications successfully mitigated the risk of **hidden UI elements** (CWE-919), indicating that no malicious or hidden interface components were detected that could be exploited to escalate privileges or confuse the user

**Table 4. Comparative Analysis of Data Storage and Privacy Compliance**

| Security Criterion | Bankak | Fawry | Okash | Sahil |
|---|---|---|---|---|
| Unsafe files deletion | ☒ | ☒ | ☒ | ☒ |
| Read/Write access to External Storage. | ☒ | ☒ | ☒ | ☒ |
| Missing copy & paste protection from EditTextfields | ☒ | ☒ | ☒ | ☒ |
| Missing protection against screenshots & screen sharing | ☒ | ☑ | ☒ | ☑ |
| No blurring for the app in the background | ☒ | ☑ | ☒ | ☒ |
| Cleartext Storage of Sensitive Information in app source code | ☑ | ☑ | ☒ | ☒ |
| Hidden elements in the user view | ☑ | ☑ | ☑ | ☑ |
| ADB Backup allowed | ☑ | ☑ | ☑ | ☑ |

## 4.2 Cryptographic Integrity

The cryptographic assessment (Table 5) identified a universal failure in secure random number generation. All evaluated applications utilize predictable Pseudo-Random Number Generator (PRNG) algorithms (CWE-330), which can lead to predictable session tokens or transaction IDs. Additionally, **Bankak** and **Okash** continue to utilize deprecated hashing algorithms (MD5/SHA-1), rendering them susceptible to collision attacks. **Fawry** and **Sahil** showed compliance in hashing but still failed the random number generation test.

**Table 5. Evaluation of Cryptographic Requirements**

| Cryptographic Criterion | Bankak | Fawry | Okash | Sahil |
|---|---|---|---|---|
| Weak Hashing Algorithms | ☒ | ☑ | ☒ | ☑ |
| Weak Random Number Generator | ☒ | ☒ | ☒ | ☒ |
| Cryptography: Predictable Initialization Vector | ☑ | ☑ | ☑ | ☑ |
| Insecure cryptographic protocols | ☑ | ☑ | ☑ | ☑ |
| Missing SQLite PRAGMA key protection | ☑ | ☑ | ☑ | ☑ |

## 4.3 Network Communication Analysis

The security of data-in-transit is paramount. As indicated in **Table 6**, **Okash** was the only application to achieve 100% compliance in this category. In contrast, **Bankak** exhibited the weakest network security posture, failing three out of four criteria.

The most critical finding is the **absence of SSL/Certificate Pinning in Bankak**, while Fawry, Okash, and Sahil successfully implemented this defense. However, unsafe TrustManager implementations were found in **Bankak, Fawry, and Sahil**, leaving them vulnerable to MitM attacks if certificate pinning is bypassed or if the attacker possesses a valid Certificate Authority (CA) certificate.

Despite the certificate pinning failures in **Bankak**, it is critical to note that all four applications correctly utilized the HTTPS scheme, avoiding the catastrophic 'Unsafe HttpHost' vulnerability where traffic is forced over unencrypted HTTP. This confirms that while the validation of certificates was flawed in some apps, the transport protocol itself was never downgraded to plaintext.

**Table 6. Results of Evaluating Network Communication Requirements.**

| Network Security Criterion | Bankak | Fawry | Okash | Sahil |
|---|---|---|---|---|
| Deprecated implementation of SSL Sockets | ☒ | ☒ | ☑ | ☑ |
| Missing Certificate Pinning | ☒ | ☑ | ☑ | ☑ |
| Unsafe TrustManager implementation | ☒ | ☒ | ☑ | ☒ |
| Unsafe HttpHost scheme for implementing https connections | ☑ | ☑ | ☑ | ☑ |

### 4.4 Platform Interaction Analysis

Platform interaction refers to how the application communicates with the Android OS and other installed applications. As shown in **Table 7**, **Fawry** demonstrated superior architectural security. However, significant vulnerabilities were identified in the remaining three applications regarding Android Component Exposure.

Specifically, **Okash** and **Sahil** improperly exported Android Services and Broadcast Receivers without adequate permission restrictions. This high-risk flaw allows malicious third-party applications installed on the same device to "bind" to the banking service or "inject" malicious intents. Furthermore, the use of raw SQL queries in **Bankak** and **Sahil**—rather than parameterized queries—introduces a risk of Local SQL Injection, where an attacker could potentially bypass authentication or dump the local application database.

Beyond component exposure, the audit confirmed that **no applications were vulnerable to WebView file access** or **unsafe Jackson deserialization**. This is a significant finding, as these vectors often allow attackers to exfiltrate local files or execute arbitrary code via malicious JSON payloads. The absence of these specific flaws suggests that while the apps struggle with component export controls, their handling of web content and data serialization remains robust.

**Table 7: Results of Evaluating Platform Interaction Requirement**

| Platform Interaction Criterion | Bankak | Fawry | Okash | Sahil |
|---|---|---|---|---|
| Raw SQL queries used for SQLite database | ☒ | ☑ | ☑ | ☒ |
| Improper Export of Android Activities | ☑ | ☑ | ☑ | ☑ |
| Improper Export of Android Services | ☑ | ☑ | ☒ | ☒ |
| Improper Export of your Android Broadcast Receiver | ☑ | ☑ | ☒ | ☑ |
| Improper Export of Android Content Providers | ☑ | ☑ | ☑ | ☑ |

| Platform Interaction Criterion | Bankak | Fawry | Okash | Sahil |
|---|---|---|---|---|
| WebView loads files from external storage | ☑ | ☑ | ☑ | ☑ |
| Downloading Files using Android Download Manager | ☑ | ☑ | ☑ | ☑ |
| Unsafe Jackson deserialization configuration | ☑ | ☑ | ☑ | ☑ |
| Fragment Injection | ☑ | ☑ | ☑ | ☑ |
| Command injection Vulnerability | ☑ | ☑ | ☑ | ☑ |

### 4.5 Code Quality Analysis

The final category of the static audit focuses on build configuration and source code maturity. As evidenced in **Table 8**, the results highlight a **universal security oversight** across the Sudanese banking sector: **Insecure Debugging Configurations**. All four evaluated MBAs (100%) were found to retain active debugging symbols and verbose execution logs within the production-release APKs.

Technically, the presence of these artifacts serves as a "developer's roadmap" for an attacker, explicitly revealing internal logic regarding API calls, data handling procedures, and endpoint structures. The failure to implement proper code obfuscation and log stripping during the build process significantly reduces the complexity of reverse engineering. By parsing these logs, a malicious actor can identify backend vulnerabilities that would otherwise remain obscured in a hardened, production-ready application.

However, the analysis also revealed positive security controls in other areas**.** Despite the retention of debugging logs, all four applications correctly set the android:debuggable flag to false, preventing the direct attachment of debuggers by malicious actors in a standard environment. Furthermore, all apps demonstrated **proper implementation of the SslErrorHandler class**, ensuring that SSL errors are not silently ignored or bypassed, which is a common vector for downgrading attacks. This indicates that while the *post-build hardening* process (stripping logs) is flawed, the *core security logic* regarding debugging permissions and SSL error handling remains intact across the sector.

**Table 8. Code Quality and Build Settings Analysis**

| Threats | Bankak | Fawry | Okash | Sahil |
|---|---|---|---|---|
| Debugging Information Provision | ☒ | ☒ | ☒ | ☒ |
| Debuggable App | ☑ | ☑ | ☑ | ☑ |
| Improper implementation of SslErrorHandler class | ☑ | ☑ | ☑ | ☑ |

### 4.6 Comparative Vulnerability Synthesis

Figure 2 provides a consolidated perspective of the applications' susceptibility to the various vulnerabilities identified across the five security domains. By synthesizing the data from Tables 4 through 8, a clear hierarchy of security posture emerges within the Sudanese mobile banking ecosystem.

As illustrated in Figure 2, **Fawry** demonstrates the highest level of security maturity, exhibiting only seven identified vulnerabilities. **Sahil** and **Okash** occupy a middle tier with 10 and 11 vulnerabilities, respectively. Conversely, **Bankak** presents the highest risk profile, with a total of 12 distinct security flaws. This summary indicates that while no application achieved full compliance with the OWASP MASVS, there is significant variance in how Sudanese financial institutions prioritize mobile application hardening.

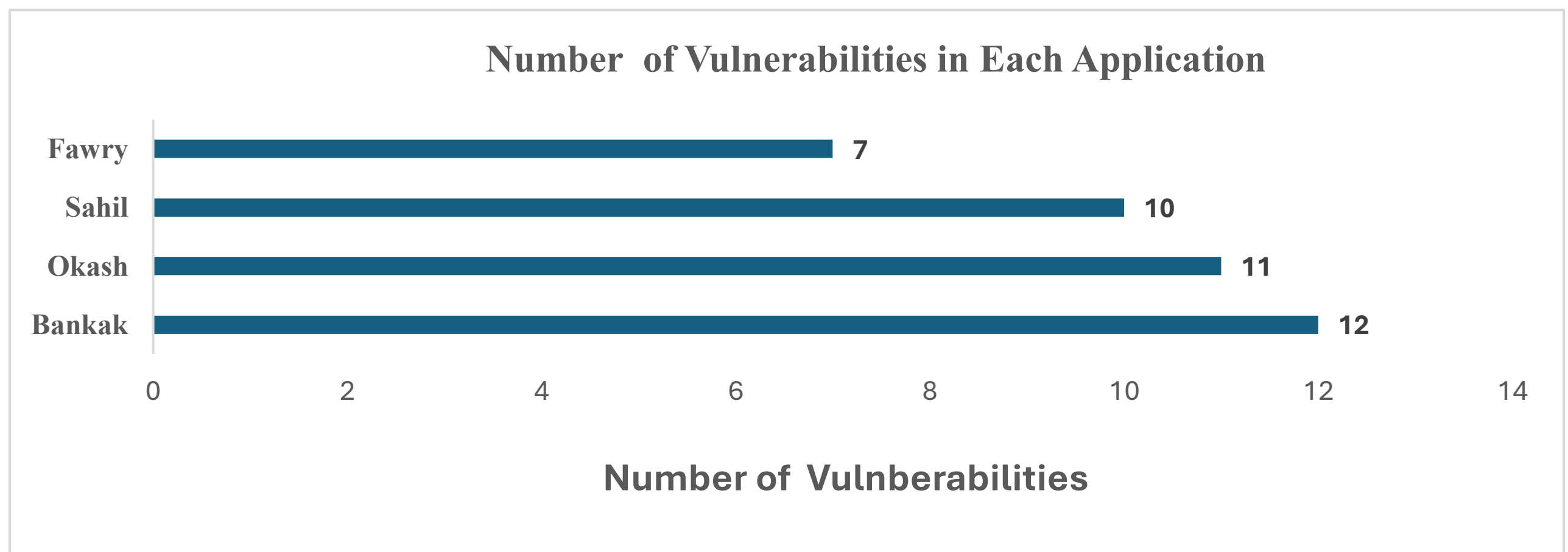

**Figure 2: Total Number of Identified Vulnerabilities per Application.**

### 4.7 Major Findings and Global Alignment

The findings reveal that Sudanese MBAs exhibit a vulnerability profile strikingly consistent with global trends in mobile financial security.

**4.7.1 Correlation with OWASP Mobile Top 10** Our analysis confirms that Sudanese banks struggle with core issues defined by the OWASP Foundation:

- **M1 (Improper Credential Usage):** Hardcoded secrets in Okash and Sahil align with global findings where approximately 50% of breaches involve compromised credentials.
- **M5 (Insecure Communication):** The lack of SSL pinning in Bankak represents a "critical" risk. International studies indicate that roughly 90% of mobile banking apps remain vulnerable to MitM attacks due to improper certificate validation.

**4.7.2 Alignment with Emerging Market Trends** Research into developing digital ecosystems (e.g., BRICS nations) highlights that "legacy device support" often leads to unsafe workarounds. Identical patterns were found in Sudan:

- **Legacy Cryptography:** The use of MD5/SHA-1 reflects a trend where banks prioritize backward compatibility over cryptographic rigor.
- **Insecure Local Storage:** Like many banking apps in West Africa, Sudanese MBAs frequently fail to use encrypted SharedPreferences, relying on the default sandbox susceptible to backup leaks.

**4.7.3 The "Usability vs. Security" Trade-off** The intentional omission of features like "background blurring" in the majority of Sudanese apps is a common global finding. Developers often disable these to reduce "user friction," unknowingly making apps susceptible to modern spyware (e.g., IdShark) that can capture financial data via screen snapshots.

# 5. Discussion

The systematic audit of Bankak, Fawry, Okash, and Sahil reveals a critical security-usability trade-off within the Sudanese fintech sector. While these applications provide essential financial inclusion, the "technical debt" in their security architecture poses a systemic risk to the national economy.

### 5.1 The Criticality of Transport Layer Failures

The most alarming finding is the absence of SSL certificate pinning in the most widely used application, **Bankak**. This oversight, combined with the unsafe TrustManager implementations found in three of the four evaluated apps, creates a fertile ground for Man-in-the-Middle (MitM) attacks. In the Sudanese context—where users frequently rely on public, unencrypted Wi-Fi networks due to infrastructure challenges—this vulnerability allows attackers to intercept and decrypt sensitive transaction data in real-time. This mirrors findings by Chen et al. (2020), who noted that global banking apps often neglect certificate pinning to reduce server-side maintenance overhead.

### 5.2 Legacy Cryptography and Data at Rest

The continued reliance on MD5 and SHA-1 for hashing sensitive metadata is a significant regression. These algorithms are considered cryptographically broken and susceptible to collision attacks. Furthermore, the failure to utilize the Android Keystore System for local data encryption means that if a device is compromised by malware or physically stolen, the application's local database is essentially "cleartext" to an attacker with root access. This aligns with the "Insecure Data Storage" trend identified by Altuwaijri and Ghouzali (2020) in emerging markets.

### 5.3 Information Leakage via Debugging Symbols

The discovery that 100% of the audited apps retained verbose debugging logs and symbols in their production-release APKs indicates a major breakdown in the Secure Software Development Life Cycle (S-SDLC). To a reverse engineer, these logs act as a "developer's roadmap," revealing internal API endpoints and logic flow. This oversight suggests that Sudanese banks may be prioritizing rapid deployment over rigorous post-build security hardening.

# 6. Recommendations

Based on the forensic findings of this study, a tiered remediation strategy is proposed. This strategy addresses immediate technical fixes for developers and long-term regulatory frameworks for the Central Bank of Sudan (CBOS).

### 6.1 Technical Remediation for Developers

To mitigate the critical vulnerabilities identified in this audit, software development teams must prioritize the following technical interventions:

- **Implement SSL Certificate Pinning:** Banks must move beyond standard HTTPS validation. Implementing certificate pinning ensures that the application only communicates with the specific server certificate, effectively neutralizing Man-in-the-Middle (MitM) attacks. This is critical for **Bankak**, the market leader, which currently lacks this protection.

- **Adopt Modern Cryptographic Standards:** Deprecated algorithms like MD5 and SHA-1 must be replaced immediately with SHA-256 or SHA-3 for hashing and AES-256 for encryption. Furthermore, developers must utilize the **Android EncryptedSharedPreferences API** and the **Android Keystore System** for local data storage to ensure that sensitive user data remains encrypted even if the device is compromised.

- **Build Hardening and Obfuscation:** The production pipeline must be integrated with code obfuscation tools such as **ProGuard** or **R8**. This process strips debugging symbols, obfuscates Smali code, and removes verbose execution logs, significantly increasing the cost and complexity of reverse engineering.

- **Enforce Secure Component Configuration:** Developers must explicitly set android:exported="false" for all Activities, Services, Broadcast Receivers, and Content Providers that do not require external access. This prevents malicious third-party apps from binding to or injecting intents into the banking application.

- **Disable ADB Backup and Screen Recording:** The android:allowBackup="false" attribute should be enforced in the manifest to prevent unauthorized local backups. Additionally, the FLAG_SECURE flag should be applied to all sensitive UI windows to prevent screenshots and screen recording by other applications or spyware.

### 6.2 Strategic Roadmap for Regulators (CBOS)

The Central Bank of Sudan (CBOS) plays a pivotal role in elevating the security posture of the national financial ecosystem. The following regulatory measures are recommended:

- **Mandatory Third-Party Auditing:** CBOS should mandate annual, independent security audits aligned with the **OWASP Mobile Application Security Verification Standard (MASVS)** as a prerequisite for maintaining a mobile banking license. These audits should include both Static (SAST) and Dynamic (DAST) testing.

- **Security-by-Design Standards:** Establish a national cybersecurity framework for fintech that explicitly penalizes the use of deprecated cryptographic standards (e.g., MD5/SHA-1) and mandates minimum security baselines for data storage and network communication.

- **Incident Response and Disclosure:** Create a centralized vulnerability disclosure program (VDP) where security researchers can responsibly report findings. This fosters a collaborative environment to identify and patch vulnerabilities before they are exploited.

- **Developer Education and Certification:** Launch a certification program for mobile banking developers focusing on secure coding practices, threat modeling, and the OWASP Mobile Top 10 to reduce the "technical debt" in the sector.

## 7. Conclusion

This study provides a comprehensive technical audit of the cybersecurity posture within Sudan's rapidly evolving fintech sector. By systematically evaluating the leading mobile banking applications—**Bankak**, **Fawry**, **Okash**, and **Sahil**—this research has moved beyond theoretical risks to identify specific, actionable vulnerabilities that threaten the local digital ecosystem.

The findings reveal a systemic lack of security maturity across the evaluated applications. Specifically, the audit identified critical failures in **transport layer integrity** (notably the absence of SSL pinning in the market leader, Bankak), **insufficient cryptographic protection** for sensitive metadata (widespread use of MD5/SHA-1), and a **universal oversight** regarding the retention of verbose debugging information in production-release APKs. These results demonstrate that while Sudan has achieved rapid digital growth, the national banking infrastructure remains significantly misaligned with international benchmarks such as the **OWASP MASVS**.

Ultimately, this research serves as a technical roadmap for Sudanese financial institutions to transition toward a "security-by-design" philosophy. By establishing a forensic-grade baseline for remediation, this audit provides the necessary evidence for regulators and developers to ensure that the continued expansion of mobile banking does not compromise the security of its users or the long-term integrity of the national financial infrastructure.

**Future Research Directions:** This study utilized Static Application Security Testing (SAST). Future research will extend this evaluation to include **Dynamic Application Security Testing (DAST)**. This will allow for the assessment of runtime protections—such as root detection, anti-tampering mechanisms, and the efficacy of SSL pinning under real-world network conditions—that cannot be fully validated through static analysis alone. Furthermore, given the rapid diversification of Sudan's financial landscape, we plan to evaluate the security posture of the burgeoning **e-wallet ecosystem**. This comparative analysis will determine if the vulnerability patterns identified in this study persist across non-banking financial platforms, ultimately contributing to a more comprehensive security framework for the nation's entire digital finance infrastructure.